\documentstyle[12pt]{article}
\font\title=cmbx10 scaled\magstep2
\newcommand{\idelt}{{\it \Delta}}

\newcommand{\M}[1]{$#1$}

\newcommand{\Ref}[1]{$^{#1}$}

\newcommand{\BM}[1]{\mbox{\boldmath{$#1$}}}

\newcommand{\Eqn}[2]{(#1$\cdot$#2)}
\makeatletter
\def\le{\mathrel{\mathpalette\gl@align<}}
\def\ge{\mathrel{\mathpalette\gl@align>}}
\def\gl@align#1#2{\lower.6ex\vbox{\baselineskip\z@skip\lineskip\z@
    \ialign{$\m@th#1\hfil##\hfil$\crcr#2\crcr=\crcr}}}
\makeatother

\begin{document}
\abovedisplayskip = 16pt
\belowdisplayskip = 16pt
\jot = 5pt
%
\setlength{\parindent}{2em}
\setlength{\baselineskip}{3.5ex}
\hfill{\vbox{\hbox{SAGA--HE--93}
             \hbox{HIROSHIMA--HUAM--31}
             \hbox{September, 1995}}}

\font\authornamel=cmcsc10 scaled\magstep1
\font\authornames=cmcsc10
\vspace*{5.0cm}
\begin{center}
{\title Scalar-Scalar Ladder Model in the Unequal-Mass Case. II  \\[1.0em]
     ---Numerical Studies of the BS Amplitudes ---} \\[2.0em]
\vspace{1.0cm}
      Ichio FUKUI and Noriaki SET\^{O}$^\dagger$ \\[1.0em]
\vspace{0.0cm}
{\sl Information Processing Center, Saga University} \\
{\sl Saga 840}\\
\vspace{0.2cm}
{\sl$^\dagger$ Department of Applied Mathematics, Faculty of Engineering} \\
{\sl Hiroshima University, Higashi-Hiroshima 739} \\[1.5em]
\vspace{0.5cm}
\end{center}
%
\begin{abstract}

   The Bethe-Salpeter amplitudes of the bound states formed
 by two scalar particles with unequal masses are analyzed
 in the massive scalar particle exchange ladder model.
 The norms of the amplitudes are calculated numerically, and it is
 confirmed that the norm vanishes for the bound state corresponding
 to the complex eigenvalue of the coupling constant.  The behaviour
 of the Bethe-Salpeter amplitudes in the momentum space is also
 investigated.
\end{abstract}

\newpage
%
\renewcommand{\thesubsection}{\S\,\arabic{subsection}.}
\begin{center} \subsection{Introduction} \end{center}

It is expected that the homogeneous Bethe-Salpeter (BS)
equation describes the bound state in quantum field theory.\Ref{\ref{Rev}}
The BS kernel appearing in the equation is usually approximated by the
contribution from the ladder graph only, and this is referred to as the
ladder model.  In this model, the BS equation can be regarded as an
eigenvalue problem for the coupling constant at a given mass of the bound
state.

    For the scalar-scalar ladder model with scalar particle exchange, if
the mass of the exchanged particle is nonzero, and if the two scalar
particles have unequal masses, the corresponding eigenvalue equation cannot
be reduced to a real form, when the invariant bound-state mass squared lies
in the physical region, that is, between zero and the two-particle
threshold.  In this case, it becomes a nontrivial problem whether the
eigenvalues of the coupling constant are real or complex.

    Naito and Nakanishi\Ref{\ref{Nai}} made an argument that all eigenvalues
would be real provided that eigenvalues had a certain analytic property.
In the same year, zur Linden\Ref{\ref{Lin}} published a paper suggesting the
reality of eigenvalues by numerical calculation, and Kaufmann\Ref{\ref{Kau}}
published a paper suggesting, on the contrary, that eigenvalues could
become complex for some mass configurations of three scalar particles and
the bound state.  Ida\Ref{\ref{Ida}} pointed out, in the next year,  that the
analyticity of eigenvalues assumed in Ref. \ref{Nai} is never satisfied if the
eigenvalue becomes non-real. Based on a rapid progress in computing
technique made some twenty years in between, a rather extensive numerical
calculation was performed to resolve these entangled situation.\Ref{\ref{Set}}
It was found that complex eigenvalues appeared in the first and second excited
states for the bound state mass squared around the pseudothreshold, and it
was also argued that the appearance could not be regarded as an artifact of
the numerical scheme employed there.

   As a continuation of Ref. \ref{Set}, which exclusively dealt with the
behaviour of eigenvalues, this paper concerns with the analysis of BS
amplitudes (eigenvectors) of the eigenvalue equation.  In the next section,
the general formalism for the BS equation is presented.  In \S\,3, the norm
of the BS amplitude is calculated numerically.  The norm of the BS
amplitude corresponding to a complex eigenvalue is found, to within the
numerical accuracy, to vanish.  In \S\,4, the BS amplitudes of some bound
states are depicted in the momentum space.  Discussions and a further
outlook are made in the final section.
\vspace{1.0em}

%
\begin{center} \subsection{BS equation and BS amplitude} \end{center}
      \hspace*{2em} The BS(Bethe-Salpeter or bound state) amplitude
\M{\phi(\BM{p},p_4)} with mass squared \M{s\,(>0)}, formed by two scalar
particles with masses \M{1+\idelt} and \M{1-\idelt} by exchanging a scalar
particle with mass $\mu$ obeys, in the ladder model, the following BS
equation (after the Wick rotation) :
\renewcommand{\theequation}{\arabic{subsection}$\cdot$\arabic{equation}}
\begin{eqnarray}
\lefteqn{\phi(\BM{p},p_4)\hspace{0.6em}=\hspace{0.6em}
                        \frac{1}{\left[(1+\idelt)^2(1-\sigma)
                        +p^2-2i(1+\idelt)\sqrt{\sigma}p_4\right]}}
                                                \nonumber \\
 & & \mbox{} \times\frac{1}{\left[(1-\idelt)^2(1-\sigma)
                        +p^2+2i(1-\idelt)\sqrt{\sigma}p_4\right]}
            \cdot \frac{\lambda}{\pi^2} \int d^4p^\prime
                   \frac{\phi(\BM{p}^\prime,p_4^\prime)}
                   {\mu^2+(p-p^\prime)^2} \; ,
\end{eqnarray}
with \M{\sigma = s/4} and \M{0 \le \idelt < 1}. We shall denote, hereafter,
the model specified above as the \M{[1+\idelt
\!\Leftarrow\!\mu\!\Rightarrow\! 1-\idelt]} model.  We regard Eq.
\Eqn{2}{1} as an eigenvalue equation for the coupling strength \M{\lambda},
which has mass dimension +2.
In this model, the two-particle threshold is
equal to 4, while the pseudothreshold is \M{4\idelt^2}.

  By introducing the polar coordinates in the four-dimensional
     momentum space \M{(\BM{p},p_4)},
\begin{eqnarray}
\lefteqn{p_4=p\cos\beta,\;\;  \BM{p}=p\sin\beta
        (\sin\theta\cos\varphi,\:
                \sin\theta\sin\varphi,\:\cos\theta)\:,}  \nonumber \\
       &  & \hspace{10em}(p \ge 0,\;\; 0 \le \beta,\;\; \theta \le \pi,\;\;
              0 \le \varphi < 2\pi)
\end{eqnarray}
we can reduce Eq. \Eqn{2}{1} to an infinite system of one-dimensional
integral equations.  Since the three-dimensional angular momentum \M{l} is
a good quantum number, we can assume that the BS amplitude has the form
\begin{eqnarray}
\phi(\BM{p},p_4) & = & \sum_{L=l}^{\infty}N_{L,l}\cdot
     (\sin\beta)^l C_{L-l}^{l+1}(\cos\beta)Y_{lm}(\theta,\varphi)
     \phi_{L,l}(p)
\end{eqnarray}
for suitable \M{l\,(=0,1,2,\ldots)} and \M{m\,(-l \le m \le l)}.  The
four-dimensional spherical harmonics are \M{(\sin\beta)^l
C_{L-l}^{l+1}(\cos\beta)Y_{lm}(\theta,\varphi)}, with associated
normalization constants \M{N_{L,l}}.

  By changing the variable \M{p} (the magnitude of the four-dimensional
momentum) to \M{z} defined by
\begin{eqnarray}
p =  \sqrt{\frac{1+z}{1-z}}\:, &  & z=\frac{p^2-1}{p^2+1}\:,
  \hspace{2em}(z:-1 \rightarrow 1 \:\:\mbox{as}\:\: p:0 \rightarrow \infty)
\end{eqnarray}
Eq. \Eqn{2}{1} is transformed into
\begin{eqnarray}
g_{L,l}(z) & = & \lambda\sum_{L^\prime=l}^{\infty}\int_{-1}^{1}
          dz^\prime\,K_{LL^\prime,l}(z,z^\prime)g_{L^\prime,l}(z^\prime).
          \hspace{5em} (L=l,l+1,\ldots)
\end{eqnarray}
The BS amplitude \M{\phi(\BM{p},p_4)} is expressed in terms of the
\M{g_{L,l}(z)\:}'s as
\begin{eqnarray}
\hspace*{-1.0em}
\phi(\BM{p},p_4) & = & \sum_{L=l}^{\infty}
p^{L}\left(\frac{2}{1+p^2}\right)^{L+3}\! i^{L-l}
         \left(\frac{|\BM{p}|}{p}\right)^l
         C_{L-l}^{l+1}\!\left(\frac{p_4}{p}\right)
         g_{L,l}\!\left(\frac{p^2-1}{p^2+1}\right)Y_{lm}(\theta,\varphi)\;.
\end{eqnarray}

     The derivation of the above equations and the explicit form of the
integral kernel matrix functions \M{K_{LL',l}(z,z')} for \M{l=0} and
\M{l=1} are given in Ref. \ref{Set}.  These matrix functions are all real,
and if the eigenvalue equation \Eqn{2}{5} admit a real eigenvalue \M{\lambda}
we can take the eigenvectors  \M{g_{L,l}(z)\;\; (L=l,l+1,\ldots)} all real.
Even in this case, however, since imaginary unit \M{i} appears on the right
hand side of Eq. \Eqn{2}{6}, the BS amplitude \M{\phi(\BM{p},p_4)} takes a
complex value in general.
\vspace{1.0em}

\begin{center} \subsection{Norm calculation} \end{center}

      In Ref. \ref{Set}, we have analyzed numerically the eigenvalue problem
\Eqn{2}{5} for the $s$-wave (\M{l=0}) case.  We have approximated the
infinite system of integral equations by a finite system of algebraic
equations: The summation over \M{L^\prime} from 0 to \M{\infty} in Eq.
\Eqn{2}{5} is truncated at some cutoff value \M{L_c}, and the integration
over \M{z^\prime} is replaced by the $N$-point Gauss-Legendre quadrature
formula.  By this procedure, we can reduce Eq. \Eqn{2}{5} to the
\M{(L_c+1)\cdot N\,}-th order matrix eigenvalue problem:
\setcounter{equation}{0}
\begin{eqnarray}
g_{L}(z_j) & = & \lambda\sum_{L^\prime=0}^{L_c}\sum_{k=1}^{N}
                K_{LL^\prime,0}(z_j,z_k)w_k\; g_{L^\prime}(z_k)\:,
                                                           \nonumber \\
                  &  & \hspace{8em} (L=0,1,\cdots,L_c,
                       \hspace{1em} j=1,2,\cdots,N)
\end{eqnarray}
where the points \M{z_1,\cdots,z_N} are the Gauss-Legendre points on the
interval \M{[-1,1]}, and \M{w_1,\cdots,w_N} are the corresponding
integration weights.  We have put \M{g_{L}(z_j) := g_{L,0}(z_j)}.

     We have found that for the \M{[1.6 \!\Leftarrow\!1.0\!\Rightarrow\!
0.4]} model (that is, for the case \M{\idelt=0.6} and \M{\mu=1.0} ), which
is the same as the case zur Linden\Ref{\ref{Lin}} and Kaufmann\Ref{\ref{Kau}}
treated,
complex eigenvalues appear in the range \M{0.25<s<2.65}, where \M{s} is the
bound-state mass squared.  Main calculations were performed for \M{N=45\;}
and \M{\;L_c=7}.  We have, in this paper, recalculate the first three
eigenvalues (denoted as \M{\lambda_0, \lambda_1 \mbox{ and } \lambda_2}) for
the case \M{s=0}(the left edge of the threshold), \M{s=0.23}(near before
the start of complex eigenvalues),  \M{s=0.27}(near after the start of
complex eigenvalues), \M{s=1.44}( the pseudothreshold), \M{s=2.62}(near
before the end of complex eigenvalues), \M{s=2.67}(near after the end of
complex eigenvalues), \M{s=3.90}(near the nonrelativistic limit), taking
\M{N=47} and \M{L_c=9}.  The result is summarized in Table I.
\vspace*{1.5em}
\begin{center}
Table I. Eigenvalues of the \M{[1.6\!\Leftarrow\!1.0\!
   \Rightarrow\!0.4]} model at several $s$'s.
 \vspace*{1em}
{\small {\renewcommand{\arraystretch}{1.3}
\begin{tabular}{r|ccccccc}  \hline\hline
$s$ & 0.00 & 0.23 & 0.27 & 1.44 & 2.62 & 2.67 & 3.90 \\ \hline
$\lambda_{0}$ & 3.448 & 3.342 & 3.323 & 2.755 & 2.113
 & 2.083 & 1.109 \\
$\lambda_{1}$ & 16.433 & 16.459 & \M{16.526-0.127i} & \M{14.887-0.734i} &
\M{13.059-0.159i} & 12.794 & 8.285 \\
$\lambda_{2}$ & 17.345 & 16.702 & \M{16.526+0.127i} & \M{14.887+0.734i} &
\M{13.059+0.159i} & 13.157 & 12.455
\end{tabular}}}
\end{center}
\vspace*{1em}

      Before calculating the norm of the BS amplitude, we shall check to
what accuracy the eigenvalue equation is satisfied.  Denoting Eq.
\Eqn{3}{1} symbolically by \M{\BM{g}=\lambda K\BM{g}}, where \M{K} is
\M{470\times 470} matrix, we calculate the relative error
\M{\parallel\BM{g}-\lambda K\BM{g}\parallel /\parallel\BM{g}\parallel}.  As
the vector norm \M{\parallel\BM{v}\parallel}, we consider two cases; the
\M{l_2}-norm ( \M{\parallel\BM{v}\parallel:=(\sum_i |v_i|^2)^{1/2}} ) and
\M{l_{\infty}}-norm ( \M{\parallel\BM{v}\parallel:=\max_i |v_i|} ).
Numerical values of the relative errors are almost same for the
\M{l_2}-norm and \M{l_{\infty}}-norm.  The relative errors corresponding to
the ground states (\M{\lambda_0}) in Table I are all less than \M{10^{-6}}.
 Rather large errors appear in the first excited state (corresponding to
\M{\lambda_{1}}) at \M{s=0} and in the first and second excited states
(corresponding to \M{\lambda_{1}} and \M{\lambda_{2}}) at \M{s=0.23}.  The
errors are, however, of order \M{10^{-4}}.  For other states, including
those having complex eigenvalues, the relative errors are less than
\M{10^{-5}}.

    As was explained in Ref. \ref{Set}, our numerical method is based on the
iteration-reduction scheme.\Ref{\ref{Fro}}  The relative error defined above
is not referred to the reduced eigenvalue problem (for the excited state),
but to the original eigenvalue problem:  We must restore the eigenvector from
the reduced eigenvector according to a certain prescription.\Ref{\ref{Fro}}
Since the eigenvalues are obtained to within 0.1\% relative error, it can be
said that our scheme can reproduce the eigenvector to the full accuracy as we
can expect.

   Having thus confirmed that the eigenfunction \M{g_{L}(z_j)} satisfies
the eigenvalue equation \Eqn{3}{1} accurately, we will calculate the norm
of the BS amplitudes corresponding to the states listed in Table I.  By the
name "norm of the BS amplitude", we mean a quantity defined by
\begin{eqnarray}
\int d^3\BM{p}\int
p_4\phi^*(\BM{p},-p_4)\left[(1+\idelt)^2(1-\sigma)+p^2-2i(1+\idelt)
\sqrt{\sigma}p_4\right]
                      \nonumber \\
\times\left[(1-\idelt)^2(1-\sigma)+p^2+2i(1-\idelt)\sqrt{\sigma}p_4\right]
\phi(\BM{p},p_4)\; .
\end{eqnarray}
This is the same as considered by Naito-Nakanishi (the left hand side of
Eq. \Eqn{2}{8} in Ref. \ref{Nai}) and by Ida (Eq. \Eqn{5}{37} in
Ref. \ref{Ida}).  This quantity can be shown to take a real value.
By making use of the expression for the BS amplitude (\M{l=0} case of
Eq. \Eqn{2}{6}), and using the recursion relation of the Gegenbauer
polynomial \M{C_{L}^{1}}, we can perform the (four-dimensional) angular
part of the integration in Eq. \Eqn{3}{2}, to the result
\vspace{-1.5em}
\begin{eqnarray}
\lefteqn{ }\nonumber \\
 &  & \int_{-1}^{1}dz
\sum_{L=0}^{\infty}\biggl\{(2-\delta_{L,0})(1-\idelt^2)
\sigma(1-z^2)+[1+z+(1-\idelt)^2(1-\sigma)(1-z)]
                 \nonumber \\
& & \times
[1+z+(1+\idelt)^2(1-\sigma)(1-z)]\biggr\}(-1)^L(1-z^2)^{L+1}\left[(\mbox{Re}
\;g_L(z))^2+(\mbox{Im}\;g_L(z))^2\right]
                 \nonumber \\
& & +\int_{-1}^{1}dz \sum_{L=0}^{\infty}4\idelt
\sqrt{\sigma}[1+z-(1-\sigma)(1-\idelt^2)(1-z)](-1)^L(1-z^2)^{L+2}
                 \nonumber \\
& & \times
\left[\mbox{Re}\;g_L(z)\mbox{Re}\;g_{L+1}(z)+\mbox{Im}\;g_L(z)\mbox{Im}\;
g_{L+1}(z)\right]
                  \nonumber \\
& & +\int_{-1}^{1}dz
\sum_{L=0}^{\infty}2(1-\idelt^2)\sigma(-1)^{L+1}(1-z^2)^{L+3}
                   \nonumber \\
& & \times
\left[\mbox{Re}\;g_L(z)\mbox{Re}\;g_{L+2}(z)+\mbox{Im}\;g_L(z)\mbox{Im}\;
g_{L+2}(z)\right]\; .
\end{eqnarray}
In the above expression, \M{\mbox{Re}\;g_L(z)} (resp.
\M{\mbox{Im}\;g_L(z)}) means the real (resp. imaginary) part of \M{g_L(z)}.
 If the corresponding eigenvalue is real, it is natural to take  \M{g_L(z)}
as a real function.  In this case \M{\mbox{Im}\;g_L(z)=0} and it can be
shown that the equality \M{\phi^*(\BM{p},-p_4)=\phi(\BM{p},p_4)} holds,
that is, the real (imaginary) part of the BS amplitude \M{\phi(\BM{p},p_4)}
is an even (odd) function of \M{p_4}.  If the eigenvalue \M{\lambda} is
complex,
\M{g_L(z)}'s acquire imaginary parts, and \M{\phi(\BM{p},p_4)} and
\M{\phi^*(\BM{p},-p_4)} become linearly independent, and these two BS
amplitudes have mutually complex conjugate eigenvalues
\M{\lambda} and \M{\lambda^*}.

    The approximate value of the norm can be obtained by putting
\M{g_L(z)=0} for \M{L} larger than the cutoff value \M{L_c(=9)}, and by
performing the integration according to the \M{N(=47)}-point Gauss-Legendre
formula.  Since the norm in Eq. \Eqn{3}{3} depends on the normalization of
\M{g_L(z)}'s, that is, an overall multiplication by a constant, it is
desirable to remove this arbitrariness.  We divide the norm by a manifestly
positive quantity obtained by replacing each summand in the integrands in
Eq. \Eqn{3}{3} by  its absolute value.  The result is
\vspace*{1em}
\begin{center}
Table II.  Norms of the BS amplitude in \M{[1.6\!\Leftarrow\!1.0\!
   \Rightarrow\!0.4]} model at several $s$'s.
\vspace*{1em}
{\small {\renewcommand{\arraystretch}{1.3}
\begin{tabular}{r|ccccccc}  \hline\hline
$s$ & 0.00 & 0.23 & 0.27 & 1.44 & 2.62 & 2.67 & 3.90 \\ \hline
$n_{0}$ & 1.000 & 0.992 & 0.990 & 0.921 & 0.761 & 0.750 & 0.189 \\
$n_{1}$ & 0.999 & 0.216 & \M{-0.186\cdot 10^{-5}} & \M{0.283\cdot 10^{-5}}
& \M{-0.324\cdot 10^{-5}} & \M{0.735\cdot 10^{-1}} & 0.238 \\
$n_{2}$ & --1.000 & --0.216 & \M{-0.186\cdot 10^{-5}} & \M{0.283\cdot
10^{-5}} & \M{-0.324\cdot 10^{-5}} & \M{-0.729\cdot 10^{-1}} & --0.280
\end{tabular}}}
\end{center}
\vspace*{1em}

     In the numerical analysis of an eigenvalue problem,  it is a general
feature that the accuracy of eigenvectors is lower than that of
eigenvalues.  Before starting our norm calculation, we expect, therefore,
that the numerical value of the norm corresponding to the zero-norm state
will be at best of order \M{10^{-3}}.  We can thus safely say that the
norms of the BS amplitudes corresponding to complex eigenvalues vanish.
Theoretically it is known that if the norm of a BS amplitude does not
vanish, this BS amplitude has a real eigenvalue.  Although the converse
statement is not true, we regard that our result amounts almost to an
analytical proof that in the
\M{[1+\idelt\!\Leftarrow\!\mu\!\Rightarrow\!1-\idelt]} model there can
appear complex eigenvalues.  We have applied our calculation procedure to
the \M{[1.0\!\Leftarrow\!1.0\!\Rightarrow\!1.0]} model (equal-mass
scalar-scalar model) and the \M{[1.6\!\Leftarrow\!0\!\Rightarrow\!0.4]}
model (unequal-mass Wick-Cutkosky model).  The norms of the first three
eigenstates at the same energies as in Table I are found to be all nonzero.
 The minimum (in absolute value) is \M{\sim 1/30}, which corresponds to the
(degenerate) first and second excited state at the pseudothreshold in the
\M{[1.6\!\Leftarrow\!0\!\Rightarrow\!0.4]} model. This value cab be
regarded well away from zero.
\vspace{1.0em}

%
\begin{center} \subsection{BS amplitude in momentum space} \end{center}

      We have verified in the preceding section that the approximate
$s$-wave eigenvector \M{g^{(k)}_{L}(z;s)}, corresponding to the $k$-th
eigenvalue \M{\lambda=\lambda_k} (\M{k=0, 1, 2}) at the bound-state mass
squared $s$, satisfies the eigenvalue equation \Eqn{3}{1} with a relative
error less than 0.001\% except for a few cases.  We can thus expect that
the corresponding approximate BS amplitude \M{\phi_k(|\BM{p}|,p_4;s)} for
\M{k=0, 1,\; \mbox{and} \; 2},
\setcounter{equation}{0}
\begin{eqnarray}
\phi_k(|\BM{p}|,p_4;s) & = & \sum_{L=0}^9
p^L\left(\frac{2}{1+p^2}\right)^{L+3}\! i^L
        C_L^1\!\left(\frac{p_4}{p}\right)
        g^{(k)}_L\!\left(\frac{p^2-1}{p^2+1};s\right)\frac{1}{2\sqrt{\pi}}
\end{eqnarray}
reflects the behaviour of the original BS amplitude in the Wick-rotated
four-dimensional momentum space rather accurately.

   As a typical example of the real eigenvalue case, Figs. 1 (a) and (b)
shows, respectively, the real and imaginary part of the BS amplitude of the
first excited state (\M{k=1}) at \M{s=0.23}.  The graphs are plotted on the
\M{p_4}-\M{|\BM{p}|} half plane, where \M{|\BM{p}|} denotes the magnitude
of the three-dimensional space part of the four-dimensional momentum.  The
amplitude is normalized so that the largest value of the magnitude of its
real part becomes 1.00.
  \vspace{1.5em}
     {\bf\large\begin{center}
         \begin{tabular}{c} \hline
          \hspace{3em} Fig. 1 (a)\hspace{3em} \\ \hline
         \end{tabular}
           \hspace{3em}
         \begin{tabular}{c} \hline
           \hspace{3em}{\bf\large Fig. 1 (b)}\hspace{3em} \\ \hline
         \end{tabular}
      \end{center}}
  \vspace{1.0em}
\hspace*{-2.0em}The expression \Eqn{4}{1} takes a form suitable for
plotting in the polar coordinates:  From the numerical calculation, we get
the data \M{g^{(k)}_L\!(z_j;s)} for
\M{L=0\;\sim\;9\;\mbox{and}\;j=-23\;\sim\;23}.  The corresponding magnitude
of the four-dimensional angular momentum \M{p= p_j:=\sqrt{(1+z_j)/(1-z_j)}}
(cf. Eq. \Eqn{2}{4}) takes the values \M{
p_{-23}=0.02532\;(z_{-23}=-0.9987),\newline
p_{-22}=0.05817\;(z_{-22}=-0.09933),\cdots,\;
p_0=1.0000\;(z_0=0.0000),\cdots,\; p_{22}=17.19 \;(z_{22}=0.9933),\;
p_{23}=39.50 \;(z_{23}=0.9987)}.  In spite of this, we have plotted the
graph in the Cartesian coordinates on the \M{p_4}-\M{|\BM{p}|} plane,
relying on the third-order interpolation-extrapolation method, in order to
make use of several useful graphic software tools.  The graph is cut if
\M{p_4} or \M{|\BM{p}|} exceeds 1.00, since the behaviour of the amplitude
is quite smooth in the region omitted.  The curves in the
\M{p_4}-\M{|\BM{p}|} plane are contour lines of the surface, and we can
read off from these curves that the real (imaginary) part of the BS
amplitude is indeed an even (odd) function of \M{p_4}, as was explained in
\S 3.

   The real and imaginary part of the BS amplitude corresponding the the
complex eigenvalue \M{\lambda_1=14.8876-0.734i} at \M{s=1.44}
(pseudothreshold) are depicted in Figs. 2 (a) and (b), respectively.
\vspace{1.5em}
      {\bf\large\begin{center}
         \begin{tabular}{c} \hline
          \hspace{3em} Fig. 2 (a)\hspace{3em} \\ \hline
         \end{tabular}
           \hspace{3em}
         \begin{tabular}{c} \hline
           \hspace{3em}{\bf\large Fig. 2 (b)}\hspace{3em} \\ \hline
         \end{tabular}
      \end{center}}
\vspace{1.0em}
\hspace*{-2.0em}In this case, the symmetry (antisymmetry) of the real
(imaginary) part of the amplitude under the \M{p_4} inversion is broken.
It seems, however, impossible to judge from Fig. 1 and Fig. 2 that the BS
norm of the former is \M{\sim 0.216} while that of the latter is \M{\sim
0.0}.

   As the last example, the absolute value of the ground state at \M{s=3.9}
is plotted in Fig. 3.
\vspace{0.5em}
      {\bf \large \begin{center}
         \begin{tabular}{c} \hline
          \hspace{3em}{\bf\large Fig. 3}\hspace{3em} \\ \hline
         \end{tabular}
       \end{center}}
\vspace{1.0em}
\hspace*{-2.0em}The support of the amplitude is concentrated near the
origin, especially in the \M{p_4} variable, as it should be in the
nonrelativistic limit.
\vspace{1.0em}

%
\begin{center} \subsection{Summary and discussion} \end{center}

      In this paper, we have analyzed numerically, as a continuation of our
previous paper,\Ref{\ref{Set}} in which the behaviour of eigenvalues of the
\M{[1.6\!\Leftarrow\!1.0\!\Rightarrow\!0.4]} model in the $s$-wave case was
clarified, the properties of the BS amplitude (eigenvector) .  We have
verified that the BS amplitude satisfies the truncated eigenvalue equation
\Eqn{3}{1} rather accurately, especially well for the amplitude of the
ground state and of the state having a complex eigenvalue.

     Based of this observation, we have calculated numerically the BS norm
of the amplitude (Eq. \Eqn{3}{2}), and found out that the norm vanishes, to
within our numerical accuracy, for the eigenvector corresponding to the
complex eigenvalue.  We can thus say with certainty that the
\M{[1+\idelt\!\Leftarrow\!\mu\!\Rightarrow\!1-\idelt]} model admits complex
eigenvalues for some configuration of \M{\idelt,\;\mu} and the bound-state
mass.   We have also depicted the BS amplitude in the momentum space for
some typical cases.

     A preliminary calculation suggests that the eigenvalues are all real
for the $p$-wave case of the \M{[1.6\!\Leftarrow\!1.0\!\Rightarrow\!0.4]}
model.  If so, it will be of some interest to see whether this is a
universal feature valid for other values of \M{\idelt} and \M{\mu}, or not.
 The appearance of complex eigenvalues will be closely connected  with the
phenomenon of anomalous threshold in the triangle Feynman graph.  These
points will be discussed in forthcoming papers.
\vspace{1.0em}

%
\begin{center} \subsubsection*{References} \end{center}
{\small
\begin{enumerate}
\item  \label{Rev}For reviews see N. Nakanishi, Prog. Theor. Phys. Suppl.
            No. 43(1969), 1. and Ed. N. Nakanishi, Prog. Theor. Phys.
            Suppl. No. 95 (1988),1.
\item  \label{Nai}S. Naito and N. Nakanishi, Prog. Theor. Phys. {\bf 42}
            (1969),402.
\item  \label{Lin}E. zur Linden, Nuovo Cim. {\bf 63A} (1969), 181.
\item  \label{Kau}W. B. Kaufmann, Phys. Rev. {\bf 187} (1969), 2051.
\item  \label{Ida}M. Ida, Prog. Theor. Phys. {\bf 43} (1970), 184.
\item  \label{Set}N. Set\^{o} and I. Fukui, Prog. Theor. Phys. {\bf 89}
            (1993),205.
\item  \label{Fro}C. E. Fr\"{o}berg, {\it Numerical Mathematics}
            (Benjamin/Cummings, California, 1985), p.189.
\end{enumerate}
}
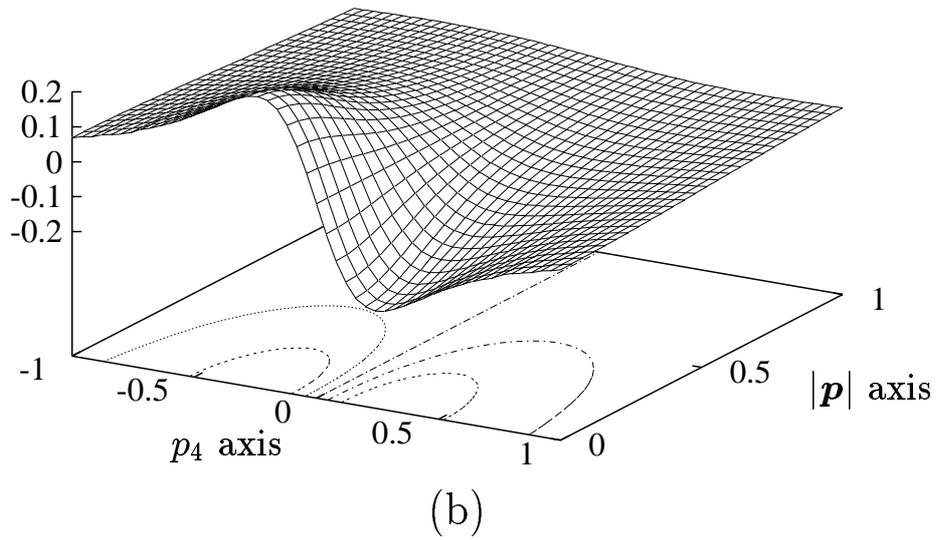
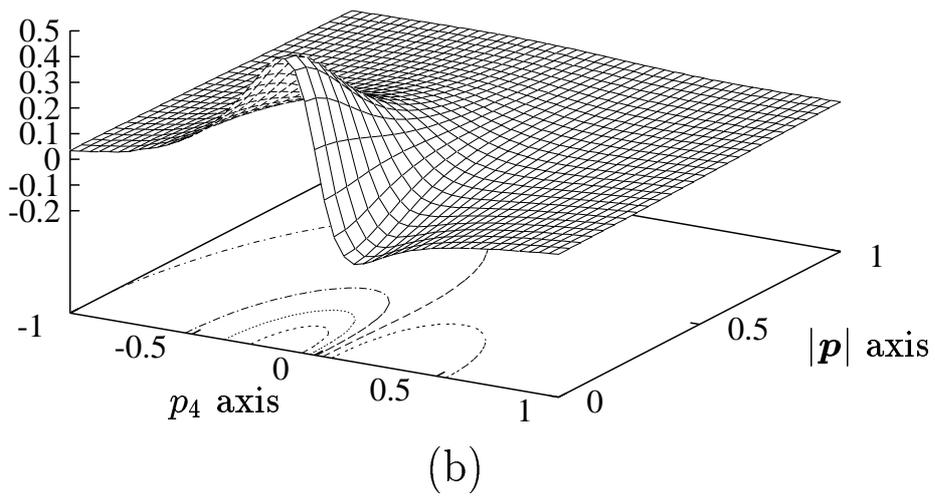
\end{document}